\documentclass[aps,prd,twocolumn,preprintnumbers,showpacs,superscriptaddress,nofootinbib]{revtex4-1}
\usepackage{graphics}  
\usepackage{epsfig}
\usepackage[usenames]{color}
 \usepackage{verbatim}

\IfFileExists{srcltx.sty}{\usepackage[active]{srcltx}}

\begin{document}
\preprint{IPMU12-0180}

\title{Moduli dark matter and the search for its decay line using
  Suzaku X-ray telescope }

\author{Alexander Kusenko}
\affiliation{Department of Physics and Astronomy, University of
California, Los Angeles, CA 90095-1547, USA}
\affiliation{Kavli Institute for the Physics and Mathematics of the Universe,
University of Tokyo, Kashiwa, Chiba 277-8568, Japan}
\author{Michael Loewenstein}
\affiliation{Department of Astronomy, University of Maryland,
College Park, MD, USA}
\affiliation{CRESST and X-ray Astrophysics Laboratory NASA/GSFC,
Greenbelt, MD, USA}
\author{Tsutomu T. Yanagida}
\affiliation{Kavli Institute for the Physics and Mathematics of the Universe,
University of Tokyo, Kashiwa, Chiba 277-8568, Japan}


\begin{abstract}
Light scalar fields called moduli arise from a variety of different
models involving supersymmetry and/or string theory; thus their
existence is a generic prediction of leading theories for physics
beyond the standard model. They also present a formidable,
long-standing problem for cosmology. We argue that an anthropic
solution to the moduli problem exists in the case of small moduli
masses, and that it automatically leads to dark matter in the form of
moduli. The recent discovery of the 125~GeV Higgs boson implies a
lower bound on the moduli mass of about a keV. This form of dark
matter is consistent with the observed properties of structure
formation, and it is amenable to detection with the help of X-ray
telescopes. We present the results of a search for such dark matter
particles using spectra extracted from the first deep X-ray
observations of the Draco and Ursa Minor dwarf spheroidal galaxies,
which are dark matter dominated systems with extreme mass-to-light
ratios and low intrinsic backgrounds. No emission line is positively
detected, and we set new constraints on the relevant new physics.
\end{abstract}

\maketitle

\section{The nature and origin of moduli}

Superstring theories, widely considered as candidates for a unified
theory of all interactions, generically predict the existence of light
scalar fields associated with the breaking of scale invariance and the
size and shape of the string compactification volume. One example is
the dilaton, a scalar field associated with the scale transformations
whose vacuum expectation value (VEV) determines the values of various
couplings in the low-energy effective field theory. String theory
compactification radii also appear in the low-energy theory as scalar
fields with some very small masses. In a number of models, the string scale
$M_{\rm s}$ is of the order of the reduced Planck mass, $M_{\rm G}=
M_{\rm P}/\sqrt{8\pi}=2.4\times10^{18}$~GeV, but, in models with a
large compactification volume, the scale of $M_s$ is suppressed by the
compactification volume and can be much lower~\cite{Witten:1996mz}.
Although one might expect that interactions of moduli with
  other fields should be suppressed by a factor that is proportional
  to some power of $M_{\rm G}$, detailed calculations can produce
  answers that are different from naive dimensional
  analyses~\cite{Svrcek:2006yi}.

Supersymmetry is a generic prediction of string theory, but, even aside
from string theory, it is a well-motivated concept in its own right. An
appealing generalization of space-time symmetries involving
non-commuting fermionic degrees of freedom in the form of a graded
algebra leads to supersymmetry and supersymmetric extensions of the
standard model. As long as supersymmetry is unbroken, its potential
has numerous flat directions with infinitely many
degenerate classical vacua. The degrees of freedom parameterizing
these flat directions are massless in the limit of unbroken
supersymmetry, but they acquire a mass from the breaking of
supersymmetry. The mass thus acquired is well below the scale at which
supersymmetry breaking occurs. For example, for $F$-type breaking, the
breaking of supersymmetry by the non-zero VEV of $|F|$ can be
communicated to the rest of the fields by gravitational interactions
suppressed by the (reduced) Planck mass $M_{\rm G}$. This gives moduli masses of
the order of $m\sim |F|/M_{\rm G}$, well below the scale of
supersymmetry breaking.  The coupling of the moduli to the rest of the fields is also
  suppressed by the higher scale.  

\section{The moduli masses}

Massless in the limit of exact supersymmetry, the moduli acquire some
masses from supersymmetry breaking. The range of moduli masses is model-dependent, and a lighter scalar with mass well
below a keV would not be a viable dark matter candidate. However, the
recent discovery of a 125~GeV Higgs boson~\cite{arXiv:1207.7214,arXiv:1207.7235} points to
the moduli masses above 1~keV, which further strengthens the
motivation for investigating moduli as a viable dark-matter candidate.

In the following section we directly address the moduli problem, and
our approach is based on a low-scale supersymmetry
breaking. Therefore, we consider the gauge-mediated supersymmetry
breaking scenario. The breaking of supersymmetry occurs via a non-zero
vacuum expectation value (VEV) of the F-component $\langle F_S \rangle
\neq 0$ of some chiral superfield $S$, whose scalar component also has
a non-zero VEV $\langle S \rangle$. Supersymmetry breaking is then
communicated to the visible sector by messengers $\Psi_i $. The
messengers couple to $S$ with couplings $\lambda_{ij}$,
\begin{equation}
 W=\lambda_{ij} S \Psi_i \bar{\Psi}_j.
\end{equation}
In what follows we suppress the indices as we discuss the generic
features of a class of models. The mass-squared matrix of the scalar
messengers takes the form
\begin{equation}
 \left ( \begin{array}{cc} |\lambda \langle S \rangle |^2 & \lambda
   \langle F_S \rangle^\dag \\ \lambda \langle F_S \rangle & |\lambda
   \langle S \rangle |^2
     \end{array} \right ).
\end{equation}
The stability condition requires that this matrix have no negative
eigenvalues, which implies
\begin{equation}
M_{\rm mess}^2 \equiv |\lambda\langle S\rangle|^2 \ \ge
\ |\lambda\langle F_S\rangle| .
\label{stability}
\end{equation}
In the visible sector, the squarks acquire masses through the gauge
interactions involving the messengers in the loops. Up to some group
theoretical factors of order one, these masses are
\begin{equation}
 m_{\rm sq} \simeq \frac{\alpha_3}{4\pi} \frac{\lambda \langle
   F_S\rangle}{M_{\rm mess}},
\label{msq}
\end{equation}
where $\alpha_3$ is the SU(3) gauge coupling constant of quantum chromodynamics (QCD).   
The recent discovery of the Higgs boson with mass 125~GeV
significantly tightens the allowed range of parameters for the squark
masses. According to the analysis
of~\cite{Okada:1990gg,Ibe:2012hu}, a mass of 125~GeV requires
that the squarks be heavier than 10~TeV. The constraint
\begin{equation}
 m_{\rm sq} > 10~{\rm TeV},
\end{equation}
combined with Eq.~(\ref{msq}), the stability constraint
Eq.~(\ref{stability}), and the requirement that $\lambda<1$ implies
the following lower limit on the supersymmetry breaking scale $
|F|\equiv |F_S|+ ({\rm other \ contributions}) > |F_S|$:
\begin{equation}
|F|\ge |F_S| > \left ( \frac{m_{\rm sq} }{10~{\rm TeV}} \right )^2
\left (10^{6}\, {\rm GeV} \right )^2.
\label{Fmin}
\end{equation}
We note that $|F|= |F_S|$ in the case of direct
mediation~\cite{hep-ph/9701286}.

This lower bound on the supersymmetry breaking scale implies the
following lower bound on the mass of the moduli:
\begin{equation}
 m_\phi = \frac{|F|}{M_{\rm G}} > 1~{\rm keV}.
\label{kevmass}
\end{equation}

Some moduli can get larger masses, for example, if they couple directly
to the gauge fields in the hidden sector, where gaugino condensation
takes place.  However, it is likely that some moduli remain massless
until supersymmetry is broken and get masses of the order of the gravitino 
mass in Eq.~(\ref{kevmass}) by gravity mediation.  We will focus on
such moduli.

Some additional constraints arise from the requirement of not
over-producing dark matter in the form of gravitinos, which have a
similar mass. For masses in the several keV range the gravitinos come
into thermal equilibrium and reach thermal abundances at temperatures
above TeV, and the total amount of dark matter would be unacceptably
large if no dilution occurred. However, gauge-mediated supersymmetry
breaking models contain candidate particles whose out-of-equilibrium
decay can produce entropy and dilute the population of dark matter to
an acceptable level~\cite{Fujii:2002fv,astro-ph/0108172}. The
required dilution factor $\Delta$ is proportional to the mass, $\Delta
\sim 100 (m_{3/2}/10 {\rm keV})$~\cite{Fujii:2002fv}. For masses
above 1~MeV, the requisite dilution of $\Delta \sim 10^4$ can be
difficult to reconcile with leptogenesis because the baryon asymmetry
is diluted by the same factor.
For masses below (but not too far below) 1~MeV the dilution
  can still be efficient, and the gravitino abundance well below the
  observed dark matter abundance. As we discuss below, the moduli
  generally tend to be over-produced, but anthropic selection
  considerations point to abundances in the range of observational
  estimates of the dark matter density. 
Since the same anthropic reasoning would not work for the gravitinos,
one can assume that the moduli have a greater abundance than the gravitinos.  
It is also possible that the contributions of the moduli and the gravitinos are comparable.  
The intriguing possibility that dark matter is composed of two
separate components with different free-streaming properties, which can
have some observable effects on the small scales, was discussed in
detail~\cite{Petraki:2008ef,Boyanovsky:2008nc,Boyanovsky:2007ba} in connection with
sterile neutrinos produced by two distinct
mechanisms~\cite{hep-ph/0602150,Kusenko:2006rh,Petraki:2007gq,Petraki:2008ef,Kusenko:2010ik}. 
In our case, there is an analogous possibility that dark matter is comprised of two components, namely, 
the moduli and the gravitinos.  

Taking into account the Higgs boson mass and the gravitino
constraints, the moduli masses of interest lie in the range $m_\phi
\simeq (1 - 10^3)$~{\rm keV}.

\section{The moduli problem}

As discussed above, a number of independent arguments based on well-motivated theories converge on 
the prediction of light scalar fields coupled very weakly to the remaining fields in the theory. 
The relevant masses range from a few eV to well above the electroweak
scale. The lighter moduli can decay through interactions suppressed by
an effective high energy scale $\Lambda_{\rm eff}$ into photons and
neutrinos. Typically, the dominant decay channel is $\phi\rightarrow
\gamma \gamma$ through a coupling
\begin{equation}
 {\cal L}_{\rm int}= \frac{1}{4 \Lambda_{\rm eff}} \, \phi \, F_{\mu\nu}F^{\mu\nu}
=\frac{b}{4M_{\rm G}} \, \phi \, F_{\mu\nu}F^{\mu\nu}, 
\end{equation}
where we have defined a parameter $b$,
\begin{equation}
 b=\frac{M_{\rm G}}{\Lambda_{\rm eff}},
\end{equation}
choosing $b$ and $\Lambda_{\rm eff}$ so as to parameterize the
coupling in a model-independent way. $\Lambda_{\rm eff}$ and the
corresponding parameter $b$ have specific meanings in the context of
specific models. For example, for string moduli, $\Lambda_{\rm eff}$
can be related to the string compactification scale and
structure. Explicit calculations yield values of $b$ ranging from
$b=\sqrt{2} \sim 1$~\cite{Dudas:1997jn} to much larger
values~\cite{Svrcek:2006yi}.

The rate of moduli radiative decay is 
\begin{equation}
 \Gamma_{\phi \rightarrow \gamma\gamma}= b^2\frac{m_\phi^3}{64\pi M_{\rm G}^2}. 
\end{equation}

The generic prediction of light scalar fields gives rise to a very
serious cosmological problem, namely the moduli problem. There is
strong and growing evidence that the universe has undergone a period
of inflation that resulted in flatness and homogeneity of the observed
universe, and that generated the observed spectrum of density
perturbations. If mass $m_\phi$ is smaller than the Hubble constant
$H$ during inflation, then the vacuum expectation value of the light
scalar field during inflation can be very large for the following
reason. One expects the energy density in each scalar field to be of
the order of $H^4$, that is $m^2\phi^2 \sim H^4$, during
inflation. Hence, the VEV $\langle \phi \rangle$, on average (i.e.,
averaged over superhorizon scales), takes a very large
value. Furthermore, the value of the ground-state VEV is not well
defined. During inflation, the minimum of the potential for the $\phi$
field can differ from its zero-temperature value due to the terms
associated with supersymmetry breaking in de~Sitter space, i.e., due
to expansion of the universe. These terms, coming from the K\"ahler
potential, act as additional mass terms $c H^2 \phi^2$, where $c$ is a
constant of order 1 that can be positive or
negative~\cite{Dine:1983ys,Dvali:1995mj}.  
These contributions overwhelm the small moduli masses and can displace
the VEV of the field by a large amount. When inflation is finished, and
reheating commences, the VEVs of the moduli fields are stuck at some
values far from the minima of their respective potentials. Eventually
the Hubble constant $H$ approaches the mass $m_\phi$; and, only then
can the field start oscillating. However, the minimum of the effective
potential, which includes the expansion-induced contributions, can
still differ from the zero temperature minimum\footnote{This stymies
  some of the attempted solutions: even if the moduli oscillations can
  be damped at some point in the expanding universe for some specific
  value of the Hubble parameter $H$, they are bound to resume when $H$
  decreases further because the minimum of the effective potential
  depends on $H$~\cite{Dvali:1995mj}.}.

A scalar field with some vanishingly small couplings to the lighter
particles oscillates until the time
\begin{equation}
 \tau_{\phi \rightarrow \gamma\gamma}=\Gamma_{\phi \rightarrow
   \gamma\gamma}^{-1}=7.6\times 10^{32} \, \left(
 \frac{1}{b}\right)^2\, \left( \frac{1 \, {\rm keV}}{m_\phi} \right)^3
      {\rm s}.
\label{eq:tau}
\end{equation}
During this time, the field's cosmological behavior is identical to
that of non-relativistic matter, with energy density declining with
scale factor $a$ as $1/a^3$. When the radiation density that scales as
$1/a^4$ drops below the energy density of moduli, the universe enters
a prolonged period of matter dominated expansion, at the end of which
moduli decay. This causes a variety of cosmological problems,
depending on the moduli mass and decay width.

If moduli decay before big-bang nucleosynthesis, their decay products
thermalize and they do not cause a conflict with observations,
assuming that the baryon asymmetry of the universe is not diluted
away. It is possible to generate the baryon asymmetry through
Affleck--Dine
baryogenesis~\cite{Affleck:1984fy,Moroi:1994rs,Dine:2003ax} so as to
compensate for dilution. Masses $m_\phi$ above $10^2$~TeV may be
considered ``safe'' in this sense. If the moduli decay during big-bang
nucleosynthesis, the entropy release may still be consistent with
successful nucleosynthesis, and the additional entropy released may
account for the small observed deviation of the effective number of
light species from the standard model
prediction~\cite{Fuller:2011qy,Menestrina:2011mz}. This case requires
further investigation.

Moduli decay after nucleosynthesis, but before recombination,
unacceptably dilutes the baryon density.
Decay after recombination also distorts the cosmic diffuse
  microwave, and contributes to the X- and gamma-ray, backgrounds in
  violation of observational constraints for some combinations of
  abundance and mass~\cite{Hashiba:1997rp,Asaka:1997rv,Asaka:1998ju},
  although cosmologically significant densities are not excluded for
  $m_{\phi}<100$~keV~\cite{Kasuya:2001tp}.

Finally, although decay times longer than the age of the universe are
consistent with the moduli being the dark matter if they can be
produced with the correct abundance, naive estimates for moduli dark
matter abundance in this case predict a much greater value than the
observed dark matter density. Indeed, let us estimate the energy
density of the universe when the modulus $\phi$ starts oscillating,
that is, when $H\sim m_\phi$, or when the temperature of the universe
is
\begin{equation}
 T_\phi\sim (90/\pi^2 g_*)^{1/4} \sqrt {M_{\rm G} m_{\phi}}.
\end{equation}
The density to entropy ratio is 
\begin{equation}
 \frac{\rho_\phi}{s} \sim \frac{m_\phi^2 \phi_0^2/2}{(2\pi^2/45) g_*
   T^3_\phi} \sim 10^5 \, {\rm GeV} \left ( \frac{m_\phi}{\rm keV}
 \right )^{1/2} \left ( \frac{\phi_0}{M_{\rm G} } \right )^{2}.
\label{rhophitos}
\end{equation}
Comparing this ratio, corrected for the entropy production, with
today's value,
\begin{equation}
 \frac{\rho_{\rm DM}}{s} = 0.2 \, \frac{\rho_{\rm c}}{s} = 3\times
 10^{-10} \, {\rm GeV},
\end{equation}
clearly reveals a drastic discrepancy.

With the help of thermal inflation~\cite{Lyth:1995hj,Lyth:1995ka}, or
by means of coupling the moduli fields to the
inflaton~\cite{Linde:1996cx,Nakayama:2011wqa} the moduli problem can
be ameliorated in some range of
masses~\cite{Hashiba:1997rp,Asaka:1997rv,Asaka:1998ju}. However,
neither of these approaches provides a complete solution.

\section{Anthropic selection}

As long as the moduli are long-lived, anthropic selection may
  be invoked to explain why moduli dark matter in the observed part of
  the universe is present in the right amount, circumventing the
  problem discussed in the previous section of the energy density in
  the oscillating moduli field taking on an excessively large
  value. 
The moduli problem is eliminated if the initial value $\phi_0$ of the
modulus field is, for whatever reason, sufficiently small that the
energy density in the moduli does not significantly exceed the
baryonic energy density. This may be compared with the requirement,
realized in the observable universe, for the dark-to-baryonic matter ratio to be
$\sim 4$ in order for conditions amenable to the the formation of
stars and planets, and, ultimately, for the existence of life, to be
attained~\cite{Tegmark:2005dy}. Therefore, one may simultaneously
attempt to explain away the moduli problem and explain the abundance
of dark matter based on the anthropic selection: life can only emerge
in those parts of the universe where the moduli energy density is
sufficiently low that the dark-to-baryonic matter ratio has the
acceptable value.

The reasoning here is analogous to that applied to the
axion~\cite{Linde:1991km}. Indeed, the abundance of string moduli
depends on the initial value $\phi_0$ (Eq.~\ref{rhophitos}) that is
determined by the VEV of the field during inflation. On average, this
VEV tends to be very large, of the order of $M_{\rm G}$. However the
large VEV may generate too much matter that is already dominant at a
temperature $T\sim 10^5$~GeV. Inflation still assures that $\Omega
=1$, so that the universe remains flat; however, in this case
structure formation proceeds very early, leading to a universe
dominated by black holes~\cite{Tegmark:1997in,Tegmark:2005dy}. If
matter comes to dominate the energy density of the universe too early,
the density perturbations can grow and become non-linear before
recombination. In this case, baryons and radiation are trapped
together inside the collapsing halos, and the baryon coupling to photons
maintains the Jeans mass at a constant value as the collapse
proceeds. As a result, the baryonic matter cannot fragment, and large
amounts of coupled baryon-radiation fluid are dragged into the
potential wells created by clumps of dark matter. As discussed
by~\cite{Tegmark:2005dy}, the end result of such a structure formation
process is a universe with supermassive black holes, photons, and
neutrinos, but without stars and planets. Such a universe is not
amenable to life. For the example in question, the matter-radiation
equality would occur at temperature
\begin{equation}
 T_{\rm eq} \sim 10^5\, {\rm GeV} \left ( \frac{m_\phi}{\rm keV} \right )^{1/2} 
\left ( \frac{\phi_0}{M_{\rm G} } \right )^2.
\end{equation}

Assuming that the baryon asymmetry escapes dilution, baryons are coupled
to radiation until $T_{\rm rec}\sim 1$~eV. Linear growth of density
perturbations may commence as early as at redshift $z\sim 10^{13}$ allowing for 
a longer epoch of growth in cosmological perturbations.  In the observed universe, 
primordial perturbations of the order of $\delta \rho/\rho\sim 10^{-5}$ are 
consistent with the data.  However, in the universe with matter domination starting as 
early as  redshift $z\sim 10^{13}$, these perturbations would have gone non-linear 
long before recombination leading to universe dominated by black holes and not stars and 
planets capable of hosting life.  In fact,  even the overdensities as small as $10^{-9}$ 
would have gone non-linear in this case.  The possibility of starting with very small  
density perturbations, $Q\equiv \delta \rho/\rho < 10^{-8}$ are disfavored by some anthropic arguments as well.  
The constraints arise from a combination of several astrophysical bounds summarized in Table~4 and Fig.~12 of Ref.~\cite{Tegmark:2005dy}, 
which shows the anthropicly allowed range as a function of $Q$ and matter density per CMB photon. 
There one finds no viable possibility for $Q < 10^{-8}$ and any matter density.  Likewise,
there is no viable scenario for any value of $Q$ for large dark matter densities.

We conclude that life is impossible in a universe where the dark
matter density exceeds the baryon density by many orders of magnitude,
and we refer the reader to \cite{Tegmark:1997in,Tegmark:2005dy} for a
 detailed discussion.

While the ``most common`` values of the initial field $\phi$ produce
universes without stars or planets, inflation does allow for some
relatively small part of the volume to have $\phi_0\ll M_{\rm G}$.
The probability of this happening at any given point in de~Sitter
space is negligible, but the {\em conditional} probability of a small
VEV, under the condition that life can evolve and produce an observer,
is actually of the order of 1 because life is highly improbable in the
large-VEV regions. Hence, anthropic selection would favor $\phi_0\sim
10^{-8} M_{\rm G}$, near the boundary of the parameter space
acceptable for the existence of life.

This motivates the search for dark matter in the form of string moduli
with masses below 1~MeV and for decay widths corresponding to
$b=1-10^3$. Cosmological data limits the lifetime of dark matter to
$\tau_{\phi\rightarrow \gamma \gamma}> 10^{24}$~s in the case of
radiative decays~\cite{Zhang:2007zzh,DeLopeAmigo:2009dc} (and
$\tau_{\phi\rightarrow \gamma \gamma}>3 \times 10^{18}$s for general
decays~\cite{DeLopeAmigo:2009dc}). Based on Eq.~(\ref{eq:tau}), this
corresponds to sub-MeV masses. In particular, one can search for the
dark matter in the form of moduli using X-ray telescopes.

\section{Search for moduli dark matter in X-rays}

The decay of moduli into two photons presents an opportunity for a possible discovery 
using X-ray telescopes.  Since this is a two-body decay, the signal is a very narrow line, whose width is 
determined by the Doppler broadening due to the motion of the dark matter particles.  The energy of a photon is (1/2) of the particle mass. 

If such a line is detected by the X-ray telescopes from compact astrophysical objects with substantial dark mater content and/or from diffuse 
radiation due to the galactic and cosmological distributions of dark matter,  the wavelength of the line will provide a measurement of the 
dark-matter particle mass.  The identity of the dark matter will still require some additional data, because moduli are not the only dark matter 
particles capable of producing an X-ray line.  For example, sterile neutrinos, which can also be dark matter~\cite{Dodelson:1993je,hep-ph/0602150,Kusenko:2006rh,Petraki:2007gq,Kusenko:2010ik}, 
are also expected to produce a line from their decay.  Therefore, a discovery of an X-ray line from sources correlated with the dark matter distribution would reveal 
the mass of the dark matter particle, but not necessarily its identity.  

In this section we will assume that moduli make up 100\% of dark matter.  If dark matter comprises two or more components, and moduli make up fraction $f<1$, the constraints 
derived for the decay width $\Gamma_\phi$ should be applied instead to the product $(f\times \Gamma_\phi)$, and those on $b$ to $(f^{1/2}\times b)$.   

\subsection{X-ray data used for this study}

As a proof of concept, we derive limits on the moduli decay rate in
the soft X-ray energy band ($\sim 0.7-10$~keV), based on observations
of the Ursa Minor and Draco dwarf spheroidal galaxies conducted using
the {\it Suzaku} X-ray Observatory.  These data were obtained as part of 
a search for dark matter in the form of sterile neutrinos~\cite{Loewenstein:2008yi,Loewenstein:2009cm,Loewenstein:2012px}.
 The advantage of using {\it
  Suzaku} data, as detailed in \cite{Loewenstein:2008yi}, lies in the low and
stable internal background of the {\it Suzaku} XIS (X-ray Imaging
Spectrometer) CCD detectors that, in combination with its sharp (by
X-ray CCD standards) energy resolution, provides a robust and
relatively sensitive capability for detecting faint spectral features
over a broad (soft X-ray) bandpass. The motivation for targeting dwarf
spheroidal galaxies is based on their proximity, high dark matter
density, and absence of competing X-ray
sources~\cite{Loewenstein:2008yi,Loewenstein:2009cm}. The Ursa Minor and Draco systems, that
have among the highest known dark matter surface
densities~\cite{Wolf:2009tu}, remain the only dwarf spheroidals observed by
{\it Suzaku} to date.

Utilizing these data, we derive limits on $\tau_{\phi\rightarrow
 \gamma \gamma}$, and hence on the moduli $b$~parameter, for $m_\phi$
in the $1.5-20$~keV mass range, as we now discuss in detail.

\subsection{Data reduction, spectral analysis, and line flux limits}

Spectral analysis of the {\it Suzaku} XIS data for Ursa Minor and
Draco were previously presented in~\cite{Loewenstein:2008yi} and~\cite{Loewenstein:2009cm}. Given
subsequent significant enhancements in calibration data, and in
reduction and analysis tools, we reprocess and reanalyze the data as
follows, and also consider the datasets jointly for the first time.

Our reprocessing proceeds along standard lines\footnote{see
  http://heasarc.gsfc.nasa.gov/docs/suzaku/analysis/abc/} and follows
that presented in~\cite{Loewenstein:2008yi}, and more recently in~\cite{Loewenstein:2012ke},
utilizing the {\it Suzaku} {\bf aepipeline} (version 1.0.1) ftool
task. Details may be found in those papers (and references
therein). Spectra are extracted from the entire fields-of-view for the
back-illuminated (BI: XIS1) and two operational front-illuminated (FI:
XIS0 and XIS3) chips, except for $4'$-radius circular regions around
the brightest point source in each XIS field of view, and $2'$-radius
circular regions around two fainter sources in the Draco
field. Spectral redistribution matrix files ({\bf rmf}) are generated
using {\tt xisrmfgen} version 2011-07-02, effective area function
files ({\bf arf}) using {\tt xissimarfgen} version 2010-11-05. Spectra
from the two FI detectors (XIS03 $\equiv$ XIS0 $+$ XIS3) are co-added
and a weighted XIS03 response function calculated from their
respective {\bf rmf} and {\bf arf} files. Finally, non-X-ray particle
background (NXB) spectra are extracted from the appropriately selected
and weighted night earth data using {\tt xisnxbgen} version
2010-08-22.  All spectral fitting is conducted using {\sc Xspec}
version
12.7.\footnote{http://heasarc.gsfc.nasa.gov/docs/xanadu/xspec/} Final
good exposure times, and total (including NXB) and ``source''
(NXB-subtracted) counts are displayed in Table 1.

\begin{table}
\begin{tabular}{lllll}\hline \hline 
{Galaxy} & {Detector} & {Time} & {Total} &
{Source} \\ \hline
UMinor & XIS03 & 138.1 & 10264 & 5745\\ 
UMinor & XIS1 & 69.05 & 8949 & 4380\\ 
Draco & XIS03 & 124.8 & 10361 & 6396\\ 
Draco & XIS1 & 62.4 & 9173 & 4636 \\ \hline \hline
\end{tabular}
\caption{Exposures Times (in ks) and $0.6-7$~keV Counts}
\end{table}

We adopt two approaches for establishing baseline models via spectral
fitting. In the first the NXB is subtracted, spectra in the
$0.6-7$~keV bandpass are grouped into bins with a minimum of 15 cts,
and best-fit models are found by minimizing the $\chi^2$
statistic. The models represent the astrophysical background, and are
composed of Cosmic X-ray Background (CXB) and Galactic X-ray
Background (GXB) components. The former is characterized as a power
law, the latter as a soft thermal plasma. For Draco a two-temperature
GXB model was required. Fits are performed jointly for Ursa Minor and
Draco (but separately for XIS03 and XIS1). However, only the CXB slope
is assumed identical in the two fields; GXB temperatures and all
normalizations (expressed in units per solid angle) are
untied. Reduced $\chi^2$ in the best fit models are 0.97 (1038 degrees
of freedom) for the XIS03, and 1.02 (918 degrees of freedom) for the
XIS1 fits. These spectra, best-fit models, and residuals are shown in
Figure 1.

\begin{figure}[ht]
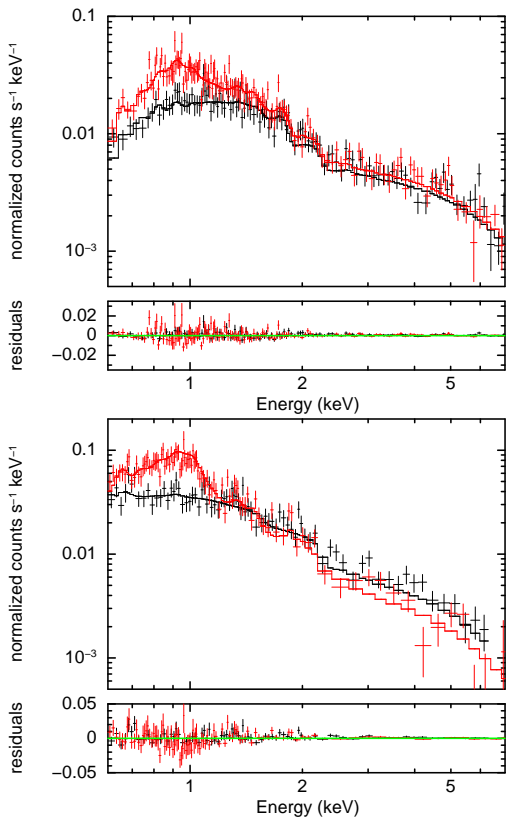

\includegraphics[scale=0.33,angle=-90]{joint_x03_0.6-7.ps}
\hfil
\includegraphics[scale=0.33,angle=-90]{joint_x1_0.6-7.ps}
\caption{\footnotesize {$0.6-7$~keV NXB-subtracted {\it Suzaku} XIS03
    ({\it left} panel, (a)) and XIS1 ({\it right} panel, (b)) spectra
    of the Ursa Minor (black) and Draco (red) Dwarf Spheroidal
    Galaxies. The best joint-fit astrophysical background baseline
    models and residuals (data-minus-model) are also shown.}
\label{fig:bs-spectra}}
\end{figure}

In the second approach, we consider the unsubtracted (total), unbinned
spectra divided into four segments chosen to separate regions with and
without strong NXB features. Overall, the spectra span the $0.7-10.5$
(XIS03) and $0.7-7$ (XIS1) keV energy ranges. Best-fit models are
found by minimizing the modification of the C-statistic~\cite{Cash:1979vz}
implemented in {\sc Xspec}. In addition to the astrophysical
background components detailed above, additional NXB power-law and
emission line components are independently included in the models in
each segment -- see~\cite{Loewenstein:2008yi} for details. The results of these
fits are shown in Figure 2.

\begin{figure}[ht]
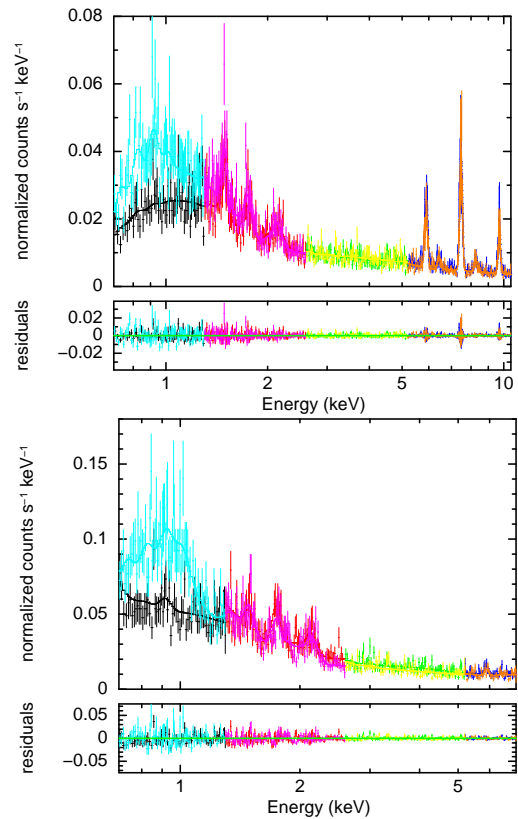

\includegraphics[scale=0.33,angle=-90]{joint_x03_nbs_new4.ps}
\hfil
\includegraphics[scale=0.33,angle=-90]{joint_x1_nbs_new4.ps}
\caption{\footnotesize {Unsubtracted {\it Suzaku} XIS03 ({\it left}
    panel, (a)) and XIS1 ({\it right} panel, (b)) spectra of the Ursa
    Minor and Draco Dwarf Spheroidal Galaxies. The best joint-fit
    baseline models, that include both non-X-ray and astrophysical
    background components, and residuals are also shown. The spectra
    are divided into segments spanning the following energies:
    $0.7-1.3$~keV (Ursa Minor: black, Draco: light blue),
    $1.3-2.6$~keV (Ursa Minor: red, Draco: magenta), $2.6-5.2$~keV
    (Ursa Minor: green, Draco: yellow), $5.2-10.5$~keV for XIS03 or
    $5.2-7$ for XIS1 (Ursa Minor: blue, Draco: orange).}
\label{fig:nbs-spectra}}
\end{figure}

Once these baseline models are established, we add an unresolved
Gaussian emission line, stepped through the full bandpass in 25~eV
steps, and derive 99\% upper confidence levels (no significant
detection was made) on the line surface brightness (that is permitted
to be negative). The joint fits are conducted such that the three
parameters of interest, in addition to the line energy, are the sum of
the Draco and Ursa Minor line surface brightnesses and their
ratio. Hence the 99\% upper confidence level corresponds to
$\Delta-statistic=11.3$.

\subsection{Dark matter surface densities}

Figure 3 shows the dynamically estimated mass profiles
from~\cite{Wolf:2009tu} and~\cite{arXiv:1009.1813}, along with other determinations in
the literature compiled by the latter. The Draco and Ursa Minor
profiles are very similar within 600~pc, consistent with an
NFW~\cite{astro-ph/9611107} profile with $M_{200}=3-30\times 10^9~\rm
M_{\odot}$\footnote{where $M_{200}$ is the mass encompassing an
  overdensity, relative to the critical density, of 200}, and a scale
radius determined by the WMAP5 mass-concentration
relation~\cite{arXiv:0805.1926}. Within apertures of $7.7'$ -- the region from
which most of the source flux originates -- and for distances of 77
(Ursa Minor) and 76 (Draco) kpc~\cite{Wolf:2009tu} the corresponding dark
matter surface density for a (tidal) truncation radius of
1.5~kpc~\cite{SanchezConde:2011ap} is $\sim 150\pm 50~{\rm M}_{\odot}~{\rm pc}^{-2}$
for each galaxy, where the cutoff is implemented using the ``$n=2$
BMO'' generalization of the NFW profile~\cite{Oguri:2011vj}. Including an
additional summed Milky Way contribution of $\sim 150~{\rm
  M}_{\odot}~{\rm pc}^{-2}$~\cite{arXiv:1109.5943}, we adopt a fiducial summed,
total line-of-sight surface mass density $\Sigma_{\rm dm}=450~\rm
M_{\odot}~{\rm pc}^{-2}$, mindful that there is an uncertainty of at
least $100~{\rm M}_{\odot}~{\rm pc}^{-2}$.

\begin{figure}[ht]
\begin{center}
\includegraphics[width=0.45\textwidth]{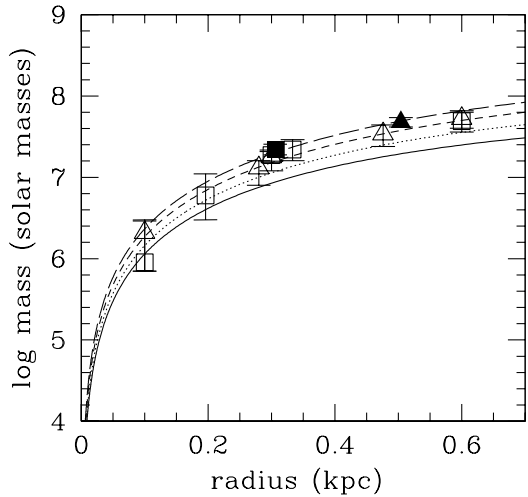}
\caption{NFW mass profiles for $M_{200}=$ 0.1, 0.3, 1.0, and 3.0
  $\times 10^{10}~\rm M_{\odot}$ (solid, dotted, short-dashed,
  long-dashed curves, respectively) compared with estimates inferred
  by~\cite{Wolf:2009tu} (filled symbols) and inferred or compiled
  by~\cite{arXiv:1009.1813} (open symbols); triangles (squares) denote Ursa Minor
  (Draco).
\label{fig:profile}}
\end{center}
\end{figure}

\subsection{Constraints on moduli dark matter}

For our estimated summed dark matter surface mass density $\Sigma_{\rm
  dm}=450~\rm M_{\odot}~{\rm pc}^{-2}$, our 99\% upper confidence
levels on emission line surface brightness yields upper limits on the
dark matter radiative decay rate over the {\it Suzaku} energy
bandpass. We compare these limits as a function of energy (in 150~eV
bins) for the joint XIS03 analysis with the predicted decay rate of
moduli dark matter for $b=10$ (equation 3) in Figure 4. Our upper
limits on $b$ as a function of $m_{\phi}$ are shown in Figure 5 for
separate emission line limits from the XIS03 and XIS1 joint spectral
analysis. Significant constraints on the moduli dark matter candidate
are obtained in the $\sim 8-20$~keV mass range from this initial
investigation. Significant improvements in sensitivity and mass range
will be realized in observations made with the {\it Astro-H}
Observatory,\footnote{http://heasarc.gsfc.nasa.gov/docs/astroh/}
scheduled for launch in 2014, with its capabilities for high energy
resolution imaging spectroscopy in soft X-rays (Soft X-ray
Spectrometer) and low background, moderate energy resolution, imaging
spectroscopy in hard X-rays (Hard X-ray Imager).

\begin{figure}
\includegraphics[width=0.45\textwidth]{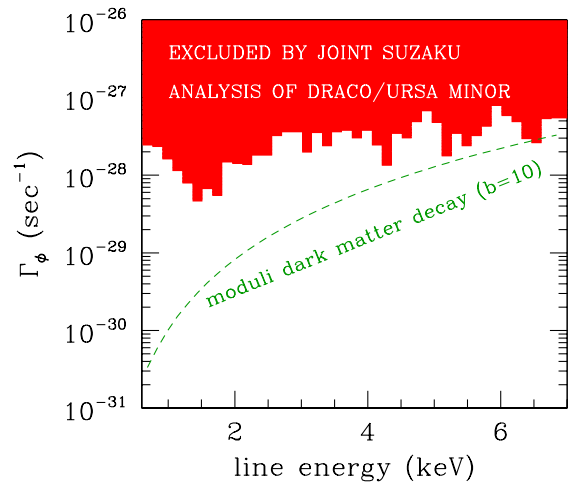}
\caption{Comparison of excluded dark matter radiative decay rate
 (shaded histogram), and predictions for moduli decay assuming
 $b=10$. The limits are based on 99\% upper confidence levels on
 line fluxes from joint analysis of Ursa Minor and Draco {\it Suzaku}
 XIS03 spectra and a summed, line-of-sight dark matter surface mass
 density $\Sigma_{\rm dm}=450~\rm M_{\odot}~{\rm pc}^{-2}$.  In the case of 
a multi-component dark matter, of which moduli make up fraction $f<1$, the same constraints 
apply to the product $(f\times \Gamma_\phi)$.
\label{fig:decay-width}}
\end{figure}

\begin{figure}
\includegraphics[width=0.45\textwidth]{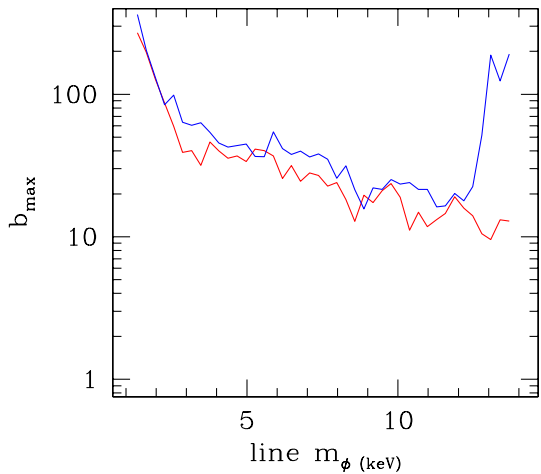}
\caption{Upper limit on the moduli $b$ parameter representing the
 (inverse of) the effective moduli scale in units of the reduced
 Planck scale, based on 99\% upper confidence levels on line fluxes
 from joint analysis of Ursa Minor and Draco {\it Suzaku} XIS03 (red)
 and XIS1 (blue) spectra and a summed dark matter surface mass
 density $\Sigma_{\rm dm}=450~\rm M_{\odot}~{\rm pc}^{-2}$. In the case of additional
dark matter components that make up $(1-f)$ of the total dark matter, the limit
is on the product $(f^{1/2}\times b)$. 
\label{fig:blimits}}
\end{figure}

\section{Conclusions}

String moduli and supersymmetry moduli produced in the early universe
may exist as a form of dark matter. Current knowledge of
supersymmetry breaking and the recent discovery of the 125~GeV Higgs
boson imply that the masses of moduli particles are above about a
keV, and the most interesting range is between 1~keV and~100 keV. In
this range, X-ray telescopes may be able to detect a line from decay
of this relic dark-matter particles.  If such a line is detected, the photon energy 
will provide information about the particle mass. 

We have presented the results of a search for such dark matter
particles using the data from the first deep X-ray observations of the
Draco and Ursa Minor dwarf spheroidal galaxies, dark matter dominated
systems with extreme mass-to-light ratios and low intrinsic
backgrounds. The absence of an emission line results in new
constraints on the relevant new physics. The limits thus obtained are sufficiently interesting so as to
  raise the prospect of a search with improved range and sensitivity
  -- and possibly a detection -- in the near future utilizing expected
  upgrades in the capabilities of X-ray observatories.

%

\section*{Acknowledgments}

The work of A.K. was supported by the DOE Grant
DE-FG03-91ER40662. A.K. appreciates the hospitality of the Aspen
Center for Physics, which is supported by the NSF Grant
No. PHY-1066293. M. L. was supported by NASA ADAP grants NNX11AD36G
and NNX11AD11G, and the Astro-H mission.

\end{document}